\documentclass[12pt]{iopart}

\usepackage{graphicx}
\usepackage{lineno}
\usepackage{url}

\newcommand{\alt}{\mathrel{\mathop{\kern 0pt \rlap
  {\raise.2ex\hbox{$<$}}}
  \lower.9ex\hbox{\kern-.190em $\sim$}}}
\newcommand{\agt}{\mathrel{\mathop{\kern 0pt \rlap
  {\raise.2ex\hbox{$>$}}}
  \lower.9ex\hbox{\kern-.190em $\sim$}}}

\newcommand{\nobb}{$0\nu\beta\beta$}
\newcommand{\twonbb}{$2\nu\beta\beta$}
\newcommand{\bb}{$\beta\beta$}

\begin{document}



\title{Micromegas readouts for double beta decay searches}




\author{S. Cebri\'{a}n$^{1}$, T. Dafni$^{1}$, E. Ferrer-Ribas$^{2}$, J. Gal\'{a}n$^{1}$, J.~A.~Garc\'{i}a$^{1}$, I.~Giomataris$^{2}$, H.~G\'{o}mez$^{1}$, D.~C.~Herrera$^{1}$, F.J.~Iguaz$^{1}$\footnote{Present address: IRFU, Centre d'\'{E}tudes de Saclay - CEA, 91191 Gif-sur-Yvette, France}, I.~G.~Irastorza$^{1}$\footnote{Corresponding author}, G. Luz\'{o}n$^{1}$, A. Rodr\'{i}guez$^{1}$, L. Segu\'{i}$^{1}$, A. Tom\'{a}s$^{1}$}

\address{$^{1}$ Laboratorio de F\'{i}sica Nuclear y Astropart\'{i}culas, Universidad de Zaragoza, 50009 Zaragoza, Spain}
\address{$^{2}$ IRFU, Centre d'\'{E}tudes de Saclay - CEA, 91191 Gif-sur-Yvette, France}

\ead{Igor.Irastorza@cern.ch}
%
%
%






\begin{abstract}

Double beta (\bb) decay experiments are one of the most active
research topics in Neutrino Physics. The measurement of the
neutrinoless mode \nobb\ could give unique information on the
neutrino mass scale and nature. The current generation of
experiments aims at detector target masses at the 100 kg scale,
while the next generation will need to go to the ton scale in
order to completely explore the inverse hierarchy models of
neutrino mass. Very good energy resolutions and ultra-low
background levels are the two main experimental requirements for a
successful experiment. The topological information of the \bb\
events offered by gaseous detectors like gas Time Projection
Chambers (TPC) could provide a very powerful tool of signal
identification and background rejection. However only recent
advances in TPC readouts may assure the competitiveness of a high
pressure gas TPCs for \bb\ searches, especially regarding the
required energy resolution. In this paper we present first results
on energy resolution with state-of-the-art \textit{microbulk}
Micromesh Gas Amplification Structure (Micromegas) using a 5.5 MeV
alpha source in high pressure pure xenon. Resolutions down to 2 \%
FWHM have been achieved for pressures up to 5 bar. These results,
together with their
recently measured radiopurity
, prove that Micromegas readouts are not only a viable option but a very competitive one for
\bb\ searches.

\end{abstract}

\vspace{3mm}                                
\begin{flushleft} \textbf{Keywords}: Neutrino, Double Beta decay, Neutrino mass, Gas detectors, Micropattern, Micromegas,
Energy resolution, Xenon
\end{flushleft}                             
%
%
%


\section{Introduction}

After the establishment of neutrino oscillations
\cite{PhysRevLett.87.071301,PhysRevLett.92.181301,Fukuda:2002pe,Fukuda:1998mi} the determination of
the neutrino mass scale is one of the most important goals in modern Neutrino Physics. If the
neutrino is a Majorana particle, the observation of the neutrinoless double beta decay would
provide an estimation of the effective Majorana electron neutrino mass\cite{Avignone:2007fu}:

\begin{equation}
m_{\beta\beta} \equiv |\sum_i U_{ei}^2 m_i | \:\:\:\:\:\:\:\:\: i
= 1,2,3
\end{equation}\label{numass}

\noindent where $U_{ei}$ are the matrix elements relating the
electron neutrino with the three mass-eigenstates and $m_i$ their
masses. This information would complement that coming from
oscillation experiments, which are only sensitive to the mass
differences of the neutrino mass-eigenstates.

The latest generation of \bb\ experiments \cite{Avignone:2007fu}
are providing sensitivities at the 0.2 -- 0.8 eV scale for the
effective neutrino mass $m_{\beta\beta}$, obtained with detector
masses at the few tens of kg scale. The current generation of setups
(being built or commissioned) aims at 100 kg scale detectors, and
will provide sensitivities down to 50-100 meV. A forthcoming
generation of experiments will need to go to the few tons scale,
and reach sensitivities down to 10-20 meV, in order to fully
explore the inverse hierarchy models of neutrino masses.

Larger detector masses are not a guarantee of better sensitivity.
These series of increasingly large experiments must come with
continuous improvements on other experimental parameters
\cite{Avignone:2007fu}. The most important requirements are on the
energy resolution and the overall experimental background in the energy region of interest (RoI), around the transition energy of the \bb\ decay, the
$Q_{\beta\beta}$ value. The latter is linked to extreme
requirements on radiopurity of the detector components, shielding
from external radiation and the availability of signal/background
discrimination techniques. The success, and eventual sensitivity,
of next generation experiments will depend on the successful
achievement of these requirements, rather than on the plain
construction of a large detector.

In general, for a next generation \bb\ experiment of about 100 kg
of target mass, background levels below $\sim10^{-3}$
c/keV/kg/year at the $Q_{\beta\beta}$ will be needed in order to
achieve its goals in a few years data taking campaign. For a detector of a few
tons to explore the inverse hierarchy region a background
level down to $\sim10^{-5}$ c/keV/kg/year could be needed. On the
other hand, the requirement on energy resolution is partially related
with the final achieved background level, as it will eventually define
the size of the RoI and therefore the background in it.
However, the need to separate the tail of the \twonbb\
distribution from the \nobb\ peak, which otherwise would account
for an irreducible background for the latter, sets an upper bound
on the energy resolution of an experiment which depends on the
intensity of the \twonbb\ signal. As an example, for $^{136}$Xe,
this argument sets a maximum accepted energy resolution of 4.5\%
FWHM for a 100 kg experiment, and a 2.5\% for a one ton experiment
\cite{Avignone:2007fu} \footnote{These values are inferred using
the current limit on the \twonbb\ in $^{136}$Xe. Since for this isotope this
decay mode has not yet been observed, the quoted values could be
larger}.

Several detection techniques are being explored in the current
generation of experiments, many of them complementary. All of them
have to rely on extreme radiopurity constraints for the innermost
detector components as well as passive and active shielding
against external radiation in order to achieve the required
background level. In fact, a large fraction of the experimental
effort of these experiments is devoted to the control and improvement of
these experimental aspects. Whether this effort will continue
yielding even lower levels of background, as needed for the next
generation of experiments, is something uncertain. Regarding energy
resolution, techniques involving semiconductors or bolometers
offer energy resolutions well better than the minimun needed, but
other techniques like scintillators or tracko-calo are already at
the limit set by the previously mentioned argument for 100 kg
experiments, and therefore their extension for larger masses will
be difficult.

The use of gas TPCs as calorimeters in
the double beta decay search of $^{136}$Xe has only recently been considered
competitive. The Gothard TPC \cite{Luscher:1998sd}
in the 90's represented a pioneering such use, although discontinued
until now due to drawbacks of the technique, like the modest energy resolution achievable.
It however demonstrated the potential of a gaseous
TPC to powerfully utilize the rich topological signature of \nobb\
events in the gas, with the characteristic two blob topology, to
further reduce background.

Since a few years several novel (or recovered) concepts and
considerations regarding the readout of gas TPCs are being put
forward that promise to overcome the limitations of conventional
TPCs and hence the prejudices against their use in \bb\ searches.
The negative ion TPC or the readout of the electroluminiscence
signal \cite{Nygren:2009zz} on one side, or the appearance and
consolidation of novel charge readout planes based on micropattern
techniques (like the Micromesh Gas Structure -- Micromegas --,
here considered), with improved energy resolution, homogeneity,
stability and scaling-up capabilities, on the other, are key
elements in these new TPC concepts. Motivated by these, at least
two collaborations (NEXT\cite{Granena:2009it} and EXO\cite{exo_tpc08}) are
proposing the use of gas TPCs for \bb\ research of $^{136}$Xe and
contemplate active work of development and prototyping.

The present research is focused on the use of Micromegas planes as
the readout of a large gaseous TPC for the \bb\ research of
$^{136}$Xe. This includes primarily the answer to the question of
whether Micromegas planes can operate satisfactorily in high
pressure pure Xenon, i.e., with sufficient gain and keeping the
good time stability and spatial homogeneity shown in other
applications (usually in Argon-based mixtures at atmospheric
pressure), and, more specifically, whether sufficiently good
energy resolutions for \bb\ are achieved at the $Q_{\beta\beta}$
energy. Operation in pure Xenon is required due to the fact that a
competitive \bb\ TPC will need to determine the absolute position
of the event along the $z$ direction (i.e. along the drift
direction), which is obtained by measuring the absolute time of
interaction, or $t_0$, given by the primary scintillation of the
Xe. The use of an additive to the xenon (quencher gas) is
therefore only allowed if it is transparent to the scintillation
signal. An scenario with such a quencher is an interesting possibility which we are also investigating. However,
in the present paper we are focused on the possibility of operating Micromegas in
pure Xenon.

In addition to the previous issues, it is also necessary to study
the radiopurity of Micromegas planes in order to know its impact
on the backgrounds, and whether it is tolerable in the extremely
radiopure environment of the inside of a \bb\ detector. First
results along this line~\cite{Cebrian:2010ta} seem to point to the
fact that state-of-the-art \emph{microbulk} Micromegas are indeed
very radiopure objects, and are very much suited for low
background applications. Finally, other less fundamental issues
must also be studied, to properly assess the feasibility of
Micromegas as a Xe \bb\ TPC, such as possible outgassing and
therefore contamination of the pure Xenon by the Micromegas
materials, or the technical issue of extracting all pixel signals
out of the high pressure vessel, via very leak-tight, low
outgassing, radiopure, high channel density feedthroughs. Although
important, none of these issues appear to be of fundamental nature
and will not be treated in this paper, which is focused on the
primary question addressed before: the
operation of Micromegas readouts in high pressure pure Xenon and
the achievable energy resolution.

Therefore, the present paper gathers the first successful results
of operation of \emph{microbulk} Micromegas planes in pure Xenon
at high pressure (up to 5 bar), including measurements of the
energy resolution at high energy using the 5.5 MeV alpha peak of
an $^{241}$Am source. It constitutes an experimental demonstration
of the possibility of using Micromegas planes in a Xenon gas TPC
for the search of \bb\ with competitive prospects. In section
\ref{MMs} a brief description of Micromegas is presented,
including a brief digression on the issue of energy resolution in
a charge amplification readout. In section \ref{csetup} the
experimental setup built specifically for the present measurements
is described in some detail. In section \ref{meas}, we present the
methodology followed, the measurements performed and the main
results obtained with pure xenon. We finish with the conclusions together with our prospects and
future plans in section \ref{conclusions}.

\section{Energy resolution in Micromegas readouts}\label{MMs}

The energy resolution in gaseous proportional counters (and by
extension in gaseous TPCs with electron
avalanche readouts) depends on many factors. Some of them can be
considered non fundamental and can in principle be overcome
(although with difficulty in practice). Examples of these are
non-uniformity of readout planes, problems of equalization of
multiple channels, ballistic deficit, attachment to
electronegative impurities of the gas or time dependencies.
The only truly fundamental effects limiting the energy resolution
are the fluctuations occurring in the number of primary
electron-ion pairs produced by the ionizing particle (and
determined by the Fano factor) as well as the fluctuations in the
number of secondary electrons produced in the avalanche initiated
by each primary electron.

The Micromegas \cite{Giomataris:1995fq, Giomataris:2004aa}
readouts make use of a metallic micromesh suspended over an
(usually pixellised) anode plane by means of insulator pillars,
defining an amplification gap of the order of 25 to 150 $\mu$m.
Electrons drifting towards the readout, go through the micromesh
holes and trigger an avalanche inside the gap, inducing detectable
signals both in the anode pixels and in the mesh. It is known
\cite{Giomataris:1998rc} that the way the amplification develops
in a Micromegas gap is such that its gain $G$ is less dependent on
geometrical factors (the gap size) or environmental ones (like the
temperature or pressure of the gas) than conventional multiwire
planes or other types of micropattern detectors based on charge
amplification. This fact allows in general for higher time
stability and spatial homogeneity in the response of Micromegas,
reducing the importance of some of the non-fundamental factors
mentioned above affecting the energy resolution. In addition, the
amplification in the Micromegas gap has less inherent statistical
fluctuations than that of multiwire proportional chambers (MWPCs), due to the faster transition from the drift field to the amplification field provided by the
micromesh \cite{Alkhazov:1970fx}.

The practical realization and operation of Micromegas detectors have
been extremely facilitated by the development of fabrication
processes which yield an all-in-one readout, in contrast to
"classical" first generation Micromegas, for which the mesh was
mechanically mounted on top of the pixelised anode. Nowadays most
of the realizations of the Micromegas concept for applications in
particle, nuclear and astroparticle physics, follow the so-called
\emph{bulk}-Micromegas type of fabrication method or, more
recently, \emph{microbulk}-Micromegas.

While the bulk Micromegas \cite{Giomataris:2004aa} uses a photo
resistive film to integrate the mesh (usually a commercial woven
mesh) and anode, being already a mature and robust manufacturing
process, the microbulk Micromegas is a more recent development
\cite{Galan:2010zz}. It allows to provide, like the bulk, all-in-one
readouts but out of double-clad kapton foils. The mesh is etched
out of one of the copper layers of the foil, and the Micromegas
gap is created by removing part of the kapton by means of
appropriate chemical baths and photolithographic techniques.
Although the fabrication technique is still under development, the
resulting readouts have very appealing features, outperforming the
bulk in several aspects. The mechanical homogeneity of the gap and
mesh geometry is superior, and in fact these Micromegas have
achieved the best energy resolutions among MPGDs with charge
amplification. Because of this, time stability of microbulk is
expected to be also better than bulk. On the other hand, they are
less robust than the bulk and for the moment the maximum size of
single readouts are of only 30 cm (the limitation coming from
equipment limitation and not fundamental). This type of readouts
are being used in the CAST experiment \cite{Galan:2010zz}.

In addition, the readout can be made extremely light and most of
the raw material is kapton and copper, two of the materials known
to be (or to achieve) the best levels of radiopurity \cite{ilias_db}.
Indeed, the first radiopurity study of Micromegas
\cite{Cebrian:2010ta} shows that current microbulk readouts contain radioactivity levels at least as low as $57\pm 25$ $\mu$Bq/cm$^2$ for  $^{40}$K,  $26\pm 14$ $\mu$Bq/cm$^2$ for $^{238}$U, $<13.9$ $\mu$Bq/cm$^2$ for $^{235}$U and $<9.3$ $\mu$Bq/cm$^2$ for $^{232}$Th.

\begin{figure*}[t]
\centering
\includegraphics[width=100mm]{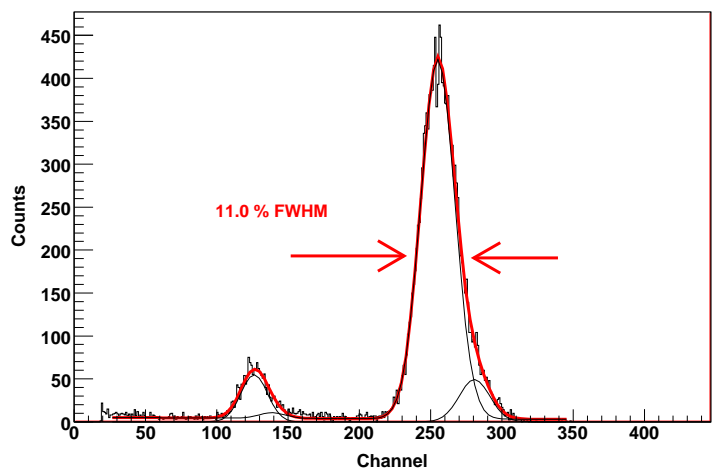}
 \caption{Typical spectrum of $^{55}$Fe with a microbulk Micromegas in an argon-isobutane mixture. The red
line is the result of a fit to 4 gaussians, while the thin black
lines are the 4 gaussians separately, corresponding to the two
x-ray emission lines of $^{55}$Fe of 5.9 and 6.5 keV, and their
corresponding escape peaks in argon.
 \label{55Fe}}
\end{figure*}

\begin{figure*}[t]
\centering
\includegraphics[width=100mm]{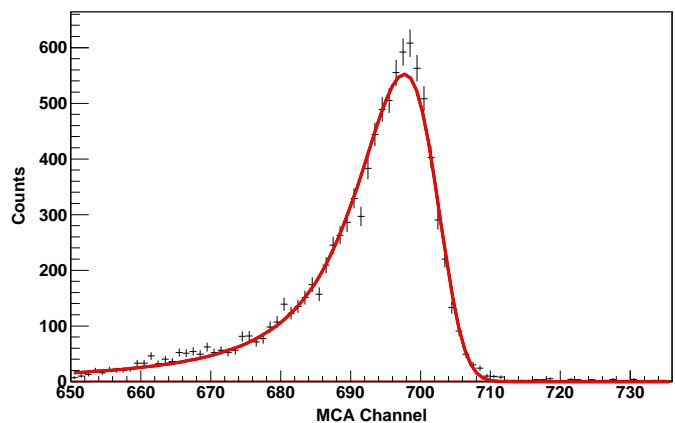}
 \caption{
Example of an alpha peak measured in \cite{Dafni:2009jb}, with the fit to a Landau function convoluted with a gaussian. The best fit values for the FWHM of
the gaussian component is 0.7\%.
 \label{alpha}}
\end{figure*}

Experimentally, resolutions of 11\% FWHM for the 5.9 keV $^{55}$Fe
peak, like the one shown in Fig. \ref{55Fe} are routinely achieved
by the current micrubulk readouts in Argon-isobutane mixtures.
Assuming a square root of energy dependency, this value would
point to energy resolutions of less then 1\% at the MeV scale,
fulfilling by far the requirements for double beta decay
applications. An experimental confirmation of this is difficult
due to the need of confining high energy events in the detection
volume. In our previous work \cite{Dafni:2009jb}, we used an
$^{241}$Am alpha source to measure energy resolutions of microbulk
readouts at high energies, obtaining values down to 1.8\% FWHM for
the 5.5 MeV alpha peak at pressures from 2 to 4 bar of argon-5\%
isobutane mixtures. Furthermore, the asymmetry of the peak pointed
to resolutions down to 0.7\% FWHM, as shown in Fig. \ref{alpha}.

The present paper aims at extending these results for the case of pure Xenon. Charge amplification in pure noble gases is problematic due to the rapid photon-driven expansion of the avalanche, which makes the detector quickly depart from the proportional amplification regime into the Geiger regime. This effect is usually avoided with the use of gas quenchers, but as argued before, this may not be possible in a Xe TPC with $t_0$ measurement based on the primary scintillation. Microbulk detectors can be built in such a way that the kapton in between mesh and anode is only removed slightly beyond the cylindrical area below every mesh hole. The avalanche is thus confined in a kapton cell, preventing the photons from expanding the avalanche far away. We speculate that this effect could work as a kind of quencher of the avalanche, and account for the modest gains indeed observed \cite{Dafni:2009jb}. This adds interest to the test of these readouts in pure noble gases. It is a fact that needs investigation by itself, and will not be explored in this paper, which is just devoted to measure the energy resolution in pure Xenon.

In \cite{Dafni:2009jb} we operated microbulks in pure Xenon, obtaining promising results, although the measurements were limited by attachment present in the system available for that work. Here a completely new setup has been built with improved tightness and outgassing conditions, in order to guarantee the purity of the Xenon.

\section{Experimental setup}\label{csetup}

\begin{figure*}[t]
\centering
\includegraphics[width=120mm]{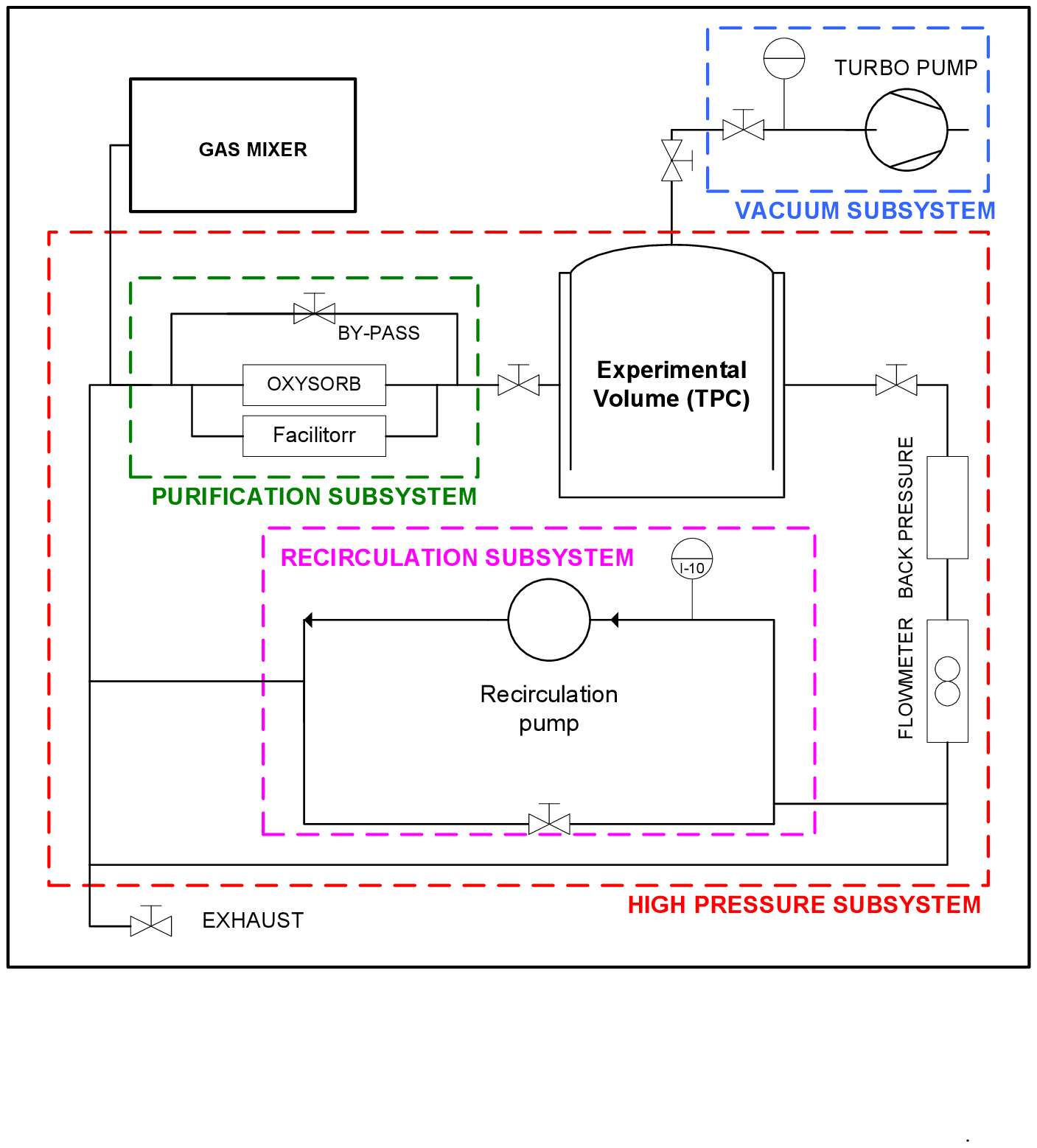}
 \caption{
Sketch of the gas filtering and recirculation setup used for measurements .
 \label{setup}}
\end{figure*}

The experimental setup was built purposely for these measurements. Its overall scheme is shown in
Fig. \ref{setup}, and consists of a high pressure vessel and a high purity gas system for
recirculation and purification. The gas system includes a gas mixer, a turbomolecular pump, a set
of gas filters, several vacuum and high pressure gauges, a high pressure recirculation pump, a
back-pressure controller, a flow meter and an exhaust line with anti-return valve.

\begin{figure}[htbp]
 \begin{minipage}{0.49\textwidth}
  \centerline{
     \includegraphics[width=0.9\textwidth, height = 1.9in]{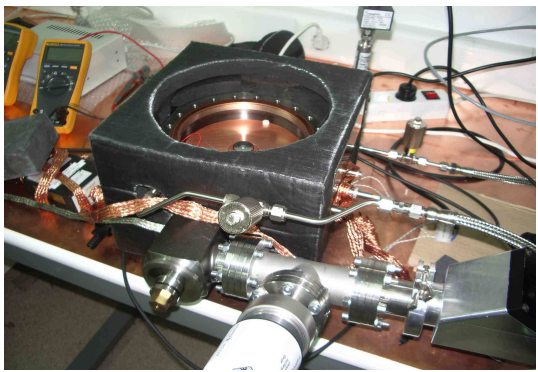}}
  \end{minipage}
  \begin{minipage}{0.49\textwidth}
  \centerline{
    \includegraphics[width=0.9\textwidth, height = 1.9in]{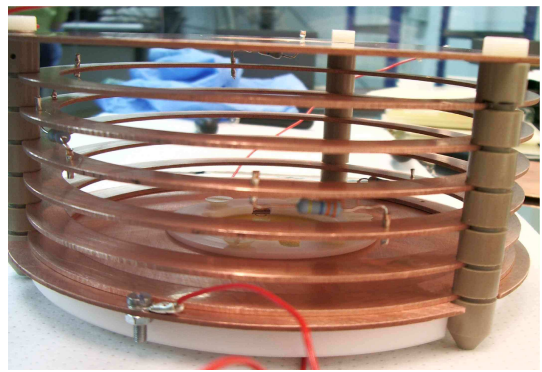}}
   \end{minipage}
  \caption{(a) Picture of the high pressure vessel, opened at the top endcap. (b) Picture of the drift and field cage structure over the Micromegas readout.}
    \label{next0-zgz}
\end{figure}

The vessel (shown in Fig. \ref{next0-zgz}) is made of stainless steel
and fulfills both ultra-high vacuum (UHV) and high pressure (up to
15 bar) requirements. Inside it has a cylindrical shape, with
10\,cm height and an internal diameter of 16\,cm. Various SHV and
BNC vacuum feed-throughs allows the electrical connections for
both the signals and voltages of the detector. They are all
implemented in CF flanges with copper joints. Specific tests were
done to assure their suitability for high pressures. A larger
15\,cm CF flange allows access to implement the
drift structure and the readout inside the vessel. Outside, the
vessel is equipped with a heat insulator blanket and four
resistors in order to bake out the system up to 110$^{\circ}$\,C.

As shown in the sketch of Fig. \ref{setup}, the vessel can be
evacuated by means of the turbo pump, through one of the CF-40
outlets. The "vacuum" part of the system (turbo pump and vacuum
gauges) is isolated from the "high pressure" part by the all-metal
valve and another vacuum valve in series. Gas can be introduced
into the vessel through the gas inlet. Gas mixtures can be
produced with the gas mixer, which is composed by 3 independent
gas lines with Bronkhorst mass flow controllers able to work at
high pressure (10 bar). Preliminary results used mixtures of argon
and isobutane, although in the present paper all results are
obtained in pure Xe. The gas is passed through a MESSER oxysorb
filter before entering the system. Once inside the vessel, the gas
can be made to recirculate along a closed loop by means of a KNF
membrane pump, and force it to continuously pass through a filter
(alternatively a MESSER oxisorb or a SAES FaciliTorr). The
recirculation flow is measured by a high precision
flowmeter and the pressure in the vessel is controlled with a
back-pressure controller.

The present system supposes a clear improvement in terms of gas
purity (initial level of vacuum, outgassing rate, capability of
filtering impurities) with respect to our previous work in
\cite{Dafni:2009jb}. A technical paper is under preparation to
describe in detail the system, the tests performed to
characterize, and different measurements performed other than the
ones of interest for the present paper \cite{iguaz_preparacion}.

The high pressure vessel was equipped with a
state-of-the-art 3.5 cm diameter circular \emph{microbulk}
Micromegas readout of 50 $\mu$m gap, with a single non-segmented
anode covering all this area. The readout laid on a metallic
support of 10 cm diameter. Apart from mechanically supporting the
Micromegas readout, it aims at extending the equipotential surface
defined by the Micromegas mesh. On top of this support the field
cage preserves a good shape of the drift electric field (i.e.
drift lines perpendicular to the Micromegas plane) all along the
conversion volume which projects onto the Micromegas surface. The
field cage includes a circular cathode 6 cm above the Micromegas
and several shaping rings (one per cm) at intermediate voltages
set by a resistor chain. The drift cathode was prepared to hold
the Americium source in its center.

The electrical connections were made in such a way that we could
power independently the drift cathode, the Micromegas mesh and the
supporting piece. Although, as explained before, the voltage of
the supporting piece is usually set the same as the one of the
mesh in order to preserve the drift field. The signal was read out
from the Micromegas mesh using a CANBERRA 2004 preamplifier, whose
output was fed into an ORTEC VT120 amplifier/shaper and
subsequently into a multichannel analyzer AMPTEK MCA-8000A for
spectra building. Alternatively, the output of the preamplifier
was digitized directly by a Tektronix TDS5054B oscilloscope and
saved into disk for further offline inspection.

\section{Measurements and results} \label{meas}

The system was cleaned before every measurement by pumping it down
to pressures below 10$^{-6}$ mbar. The purity of the Xe gas used
in the measurements here presented was of grade 6.0 (99.9999\%)
provided by Praxair. The actual purity of the gas during the
measurements, determined also by the outgassing of materials
inside the vessel and the effect of the filters was proven to be
sufficient as no evidence of attachment was seen with moderate drift fields.

%
%

The $^{241}$Am source used for the measurements consisted on a metallic circular substrate of 8
mm diameter with the Americium deposited on its center, in an
approximate circular region of about 5 mm diameter. The source is
not sealed, that is, no material is present on top of the
americium that could stop the alphas. It was installed inside the
vessel, attached to the center of the drift cathode, and in
electrical contact with it, facing the center of the Micromegas
readout. The intensity of the source was $\sim 10$ kBq.

\begin{figure}[htb!]
\centering
\includegraphics[width=120mm]{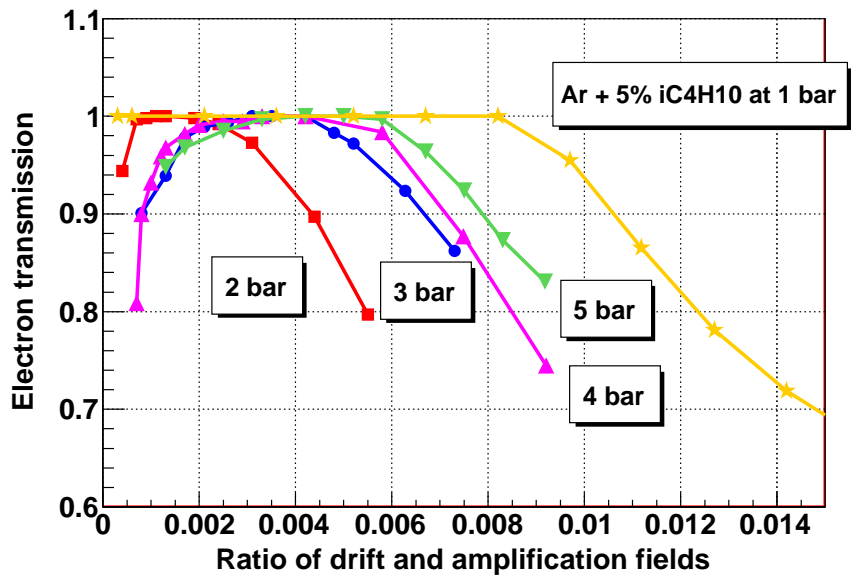}
\caption{Dependence of the alpha peak position on the ratio of drift and amplification fields ({\it electron transmission curve}) at 2 (red squares), 3 (blue circles), 4 (magenta triangles) and 5 bar (inverted green triangles). As a comparison, the same curve for Ar + 5\% iC$_4$H$_{10}$ at 1 bar (yellow stars) is included. Values are normalized to the maximum of each series.}
\label{fig:transxenon}
\end{figure}

All data here presented were taken in short runs for which sealed-mode operation was sufficient, i. e. the vessel was filled up to the desired pressure with the output outlet closed, and the gas remained static in the vessel during the run (only one pass through the input filter). In this way we operated the system at four different gas pressures: 2, 3, 4 and 5 bar. The first one of 2 bar was the minimum pressure where the projection of alpha tracks on the readout is fully contained. The maximum of 5 bar was a limit imposed indirectly by the relatively intense source available for the tests. Although higher gains were possible with the source collimated (and therefore less intense) \cite{Dafni:2009jb}, the present results (taken necessarily with the source uncollimated) were limited to gains of only about 10-20, due to the amount of total charge density generated in the Micromegas gap (related with the Raether limit). Due to this limitation, at pressures above 5 bar, the minimum drift fields needed to overcome attachment corresponded to field ratios out of the plateau of Fig. \ref{fig:transxenon}, suffering from loss of electron transmission through the mesh. Current work to extend the present results to higher pressures is ongoing, including the use of a weaker alpha source, and the use of the recirculation system to reduce the effects of attachment at low fields.

For each of the indicated pressures, the data-taking includes the systematic variation of the mesh and drift voltages. The mesh voltage is typically varied in the range from -100 V to -600 V (the Micromegas anode is kept at ground), depending on pressure, corresponding to amplification fields from 40 to 120 kV/cm. The drift voltage is varied from -250 V up to about -4000 V corresponding to drift fields from 15 V/cm to 770 V/cm. For each case, quick spectra with the MCA are registered, as well as a number of (at least $5 \times 10^4$) pulse waveforms with the oscilloscope. These pulse shapes are analysed (PSA) offline, to extract pulse characteristics like risetime and amplitude. Spectra shown later on are always based on the pulses amplitude given by the PSA.

The evolution of the peak position versus the drift/amplification field ratio gives the electron transmission curve, shown in Fig.~\ref{fig:transxenon} for each of the pure Xe pressures, and compared with a typical such curve taken in argon-isobutane at atmospheric pressure. The characteristic plateau of this curve, which usually extends to a field ratio of $\sim$0.01 for this type of readout~\cite{Dafni:2009jb}, is considerably shorter for pure Xe, something than can be qualitatively explained by the larger electron diffusion in pure Xe~\cite{nikolopoulos_rd51}. The shortening of the plateau at low drift fields for the highest pressures is due to attachment.

\begin{table}[htb!]
\begin{center}
$$
\begin{array}{ccc}
\hline
{\mbox{\bf Pressure (bar)}}&\multicolumn{2}{c}{\mbox{\bf Energy resolution (\% FWHM)}}\\
{\mbox{}}&{\mbox{raw data}}&{\mbox{risetime cut}}\\
\hline
{2.0}&{3.4}&{2.2}\\
{3.0}&{2.7}&{2.2}\\
{4.0}&{2.5}&{1.8}\\
{5.0}&{3.2}&{2.4}\\
\hline
\end{array}
$$
\end{center}
\caption{Energy resolutions measured (both from the raw data and the data after the selection on risetimes explained in the text) in pure xenon for pressures between 2 and 5 bar.}
\label{tab:eresvalues}
\end{table}

\begin{figure}[htb!]
\centering
\includegraphics[width=70mm]{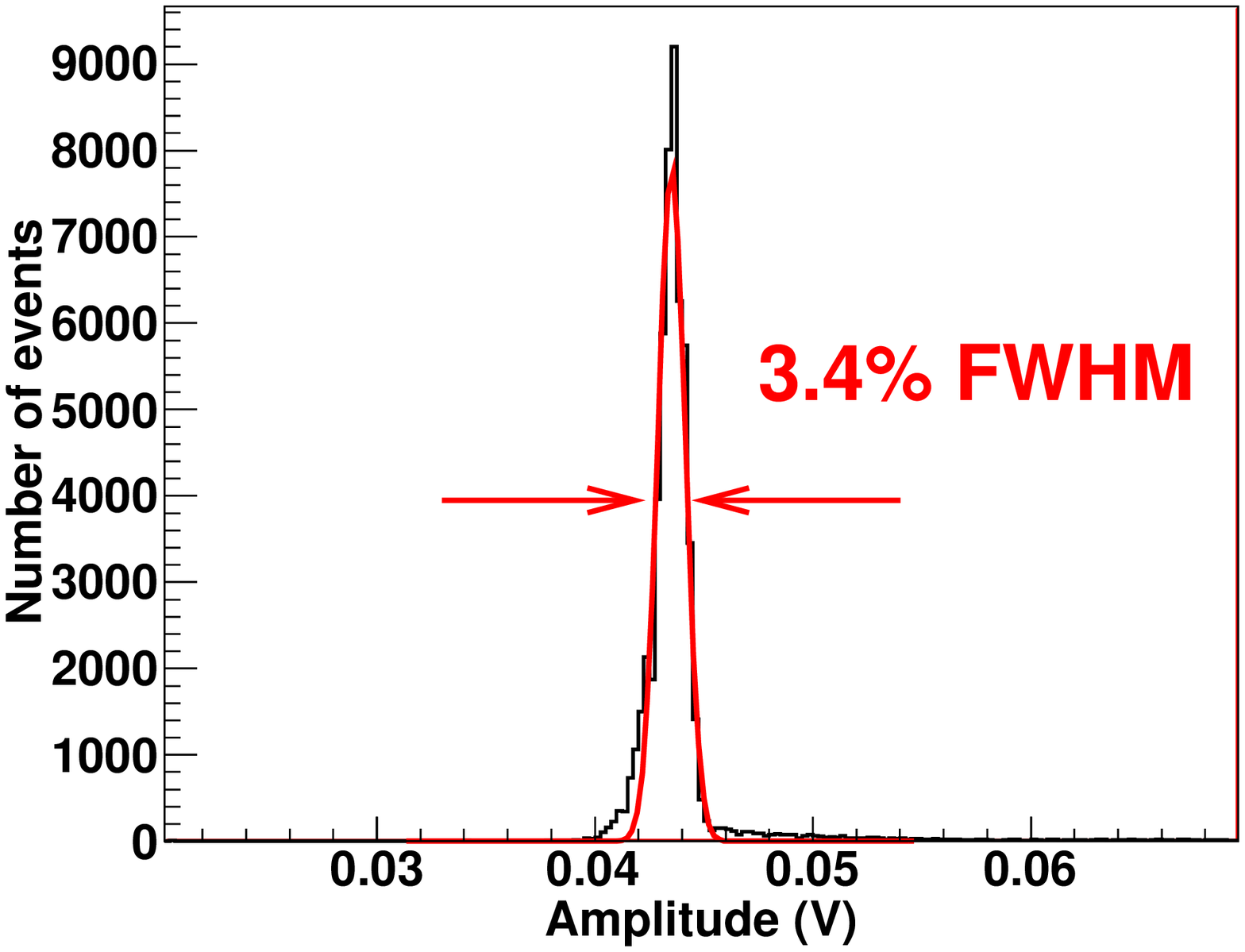}
\includegraphics[width=70mm]{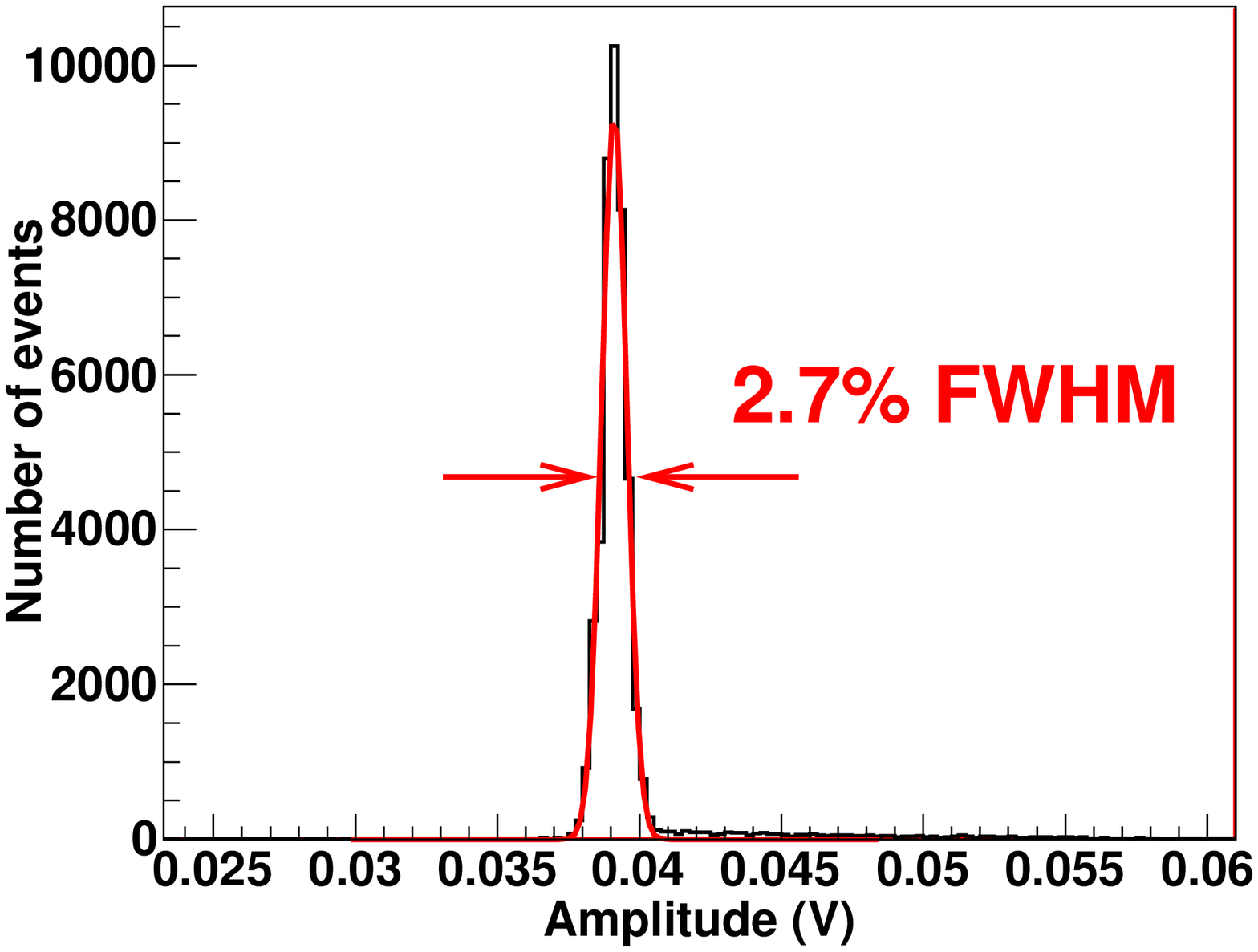}
\includegraphics[width=70mm]{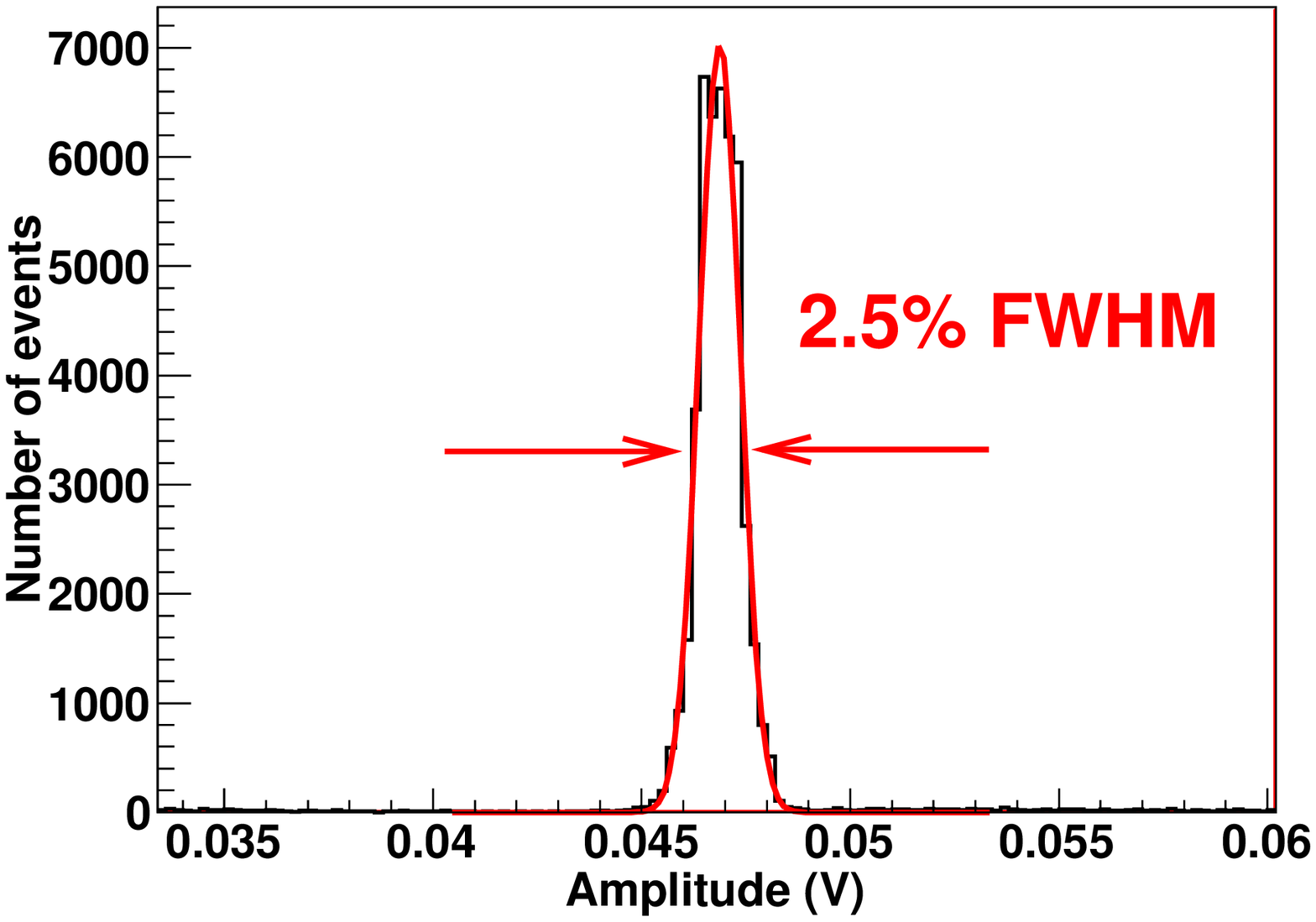}
\includegraphics[width=70mm]{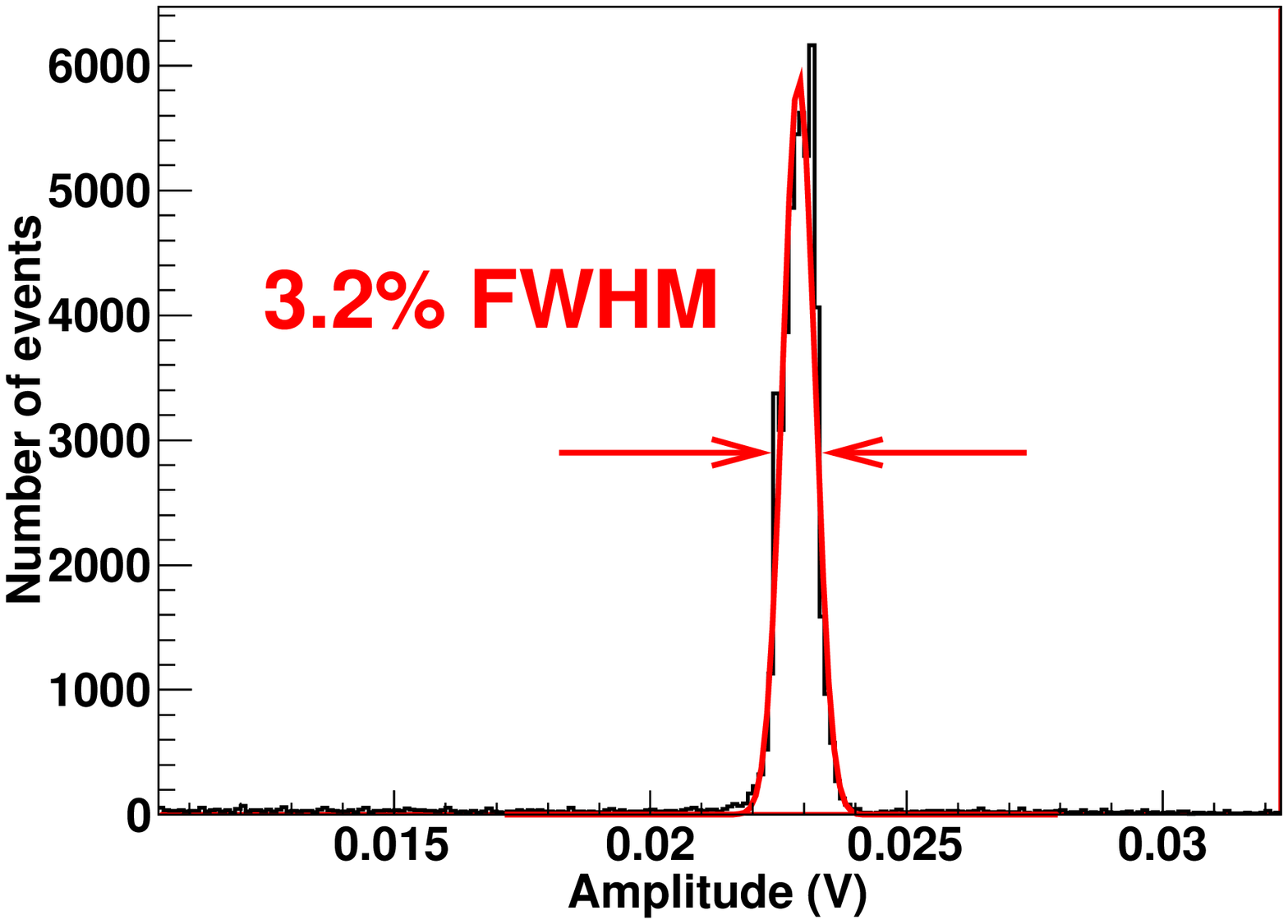}
\caption{Energy spectra of the $^{241}$Am alpha peak measured in the setup at 2 (top left), 3 (top right), 4 (bottom left) and 5 bar (bottom right). The red line and the value for the energy resolution are the result of the fit to a Gaussian function.}
\label{ref:spectra}
\end{figure}

The energy resolution of the alpha peak is extracted from the raw spectra by means of gaussian fits, like the ones shown in Fig.~\ref{ref:spectra}. The values obtained are reproduced in all over a range of values for the drift and amplification fields roughly corresponding to the plateaus of Fig.~\ref{fig:transxenon}, and are listed in table \ref{tab:eresvalues}. In general, energy resolutions at the 3~\% FWHM level have been achieved, with a best value of 2.5~\% FWHM for the 4 bar data series. These results improve considerably our previous preliminary results~\cite{Dafni:2009jb}, something that we attribute to the improved gas purity conditions in our current setup.

\begin{figure}[htb!]
\centering
\includegraphics[width=100mm]{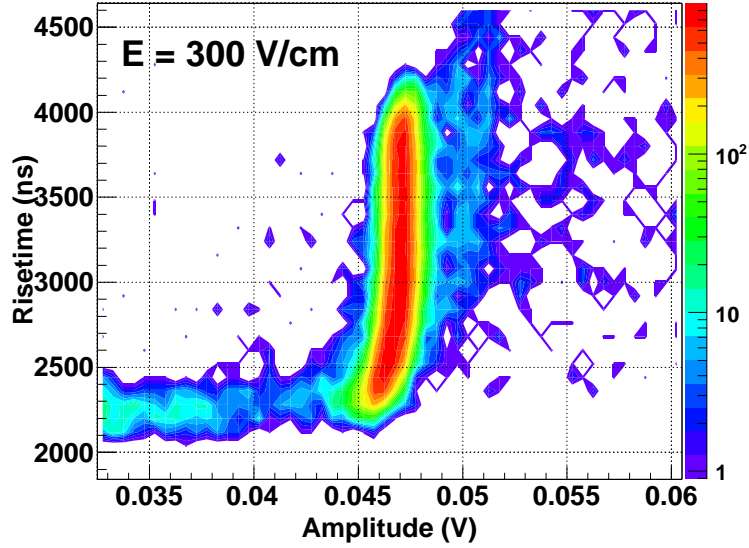}
\caption{Risetime-amplitude distributions for pure xenon at 4 bar
obtained at drift field of 300 V/cm. The gradation of color
indicates the number of events in logarithmic scale. }
\label{fig:correlation}
\end{figure}

By inspection of amplitude-risetime plots like the one shown in
Fig.~\ref{fig:correlation}, we observe, unlike in
\cite{Dafni:2009jb}, no effect of attachment (with the possible
exception of the 5 bar series, in which a slight correlation is
present at low fields). There is, however, a remaining correlation
not linked with attachment, and affecting the events with the
shortest risetime, which show amplitudes slightly below the center
of the population (clearly seen in Fig.~\ref{fig:correlation}).
These events correspond to alpha tracks with the highest angle
with the vertical, i. e. alpha tracks almost horizontal and
therefore very close to the cathode in all their trajectory. It
could be that part of the ionization of those events is deposited
in the source itself, scattered back to the cathode, or in any
case lost by some edge effect induced by the proximity of the
cathode. The contribution of these events to the energy resolution
can be removed by performing selection of events by risetime
before the gaussian fit. In this way, the values of the energy
resolution obtained, listed in the last column of
table~\ref{tab:eresvalues}, improve to values around 2\% FWHM,
with a best value of 1.8\% FWHM for the 4 bar series.

The extrapolation of this result to the $Q_{\beta\beta}$ energy of the \bb\ event depends on whether alphas suffer from any ionization quenching with respect to electrons. The authors of \cite{Luscher:1998sd} report a factor as large as 6.5 in the relative ionization of electrons with respect to alphas in a xenon-CF$_4$ mixture, although we did not measure any difference in our study in argon \cite{Dafni:2009jb}. If the value of 6.5 is true for pure Xe, our result would point to energy resolution down to 1\% or below for a $Q_{\beta\beta}$ electron energy. In the most conservative side, if no difference in ionization yield is assumed between alphas and electrons, and the resolution just scales with the square root of energy (disregarding constant energy-independent additive term to energy resolution), the expected value would be around 3\% FWHM. Any intermediate situation is possible. In any case, the present result is already sufficiently good to qualify Micromegas readouts for \nobb\ experiments using the high pressure gas Xe TPC technique.

\section{Conclusions and prospects}\label{conclusions}

We have presented measurements of energy resolution of 5.5 MeV $^{241}$Am alphas, with a Micromegas microbulk readout at high pressure pure xenon, using a new experimental setup built for this work. Energy resolutions around 3~\% FWHM are obtained for pressures of 2 to 5 bar, being the best one of 2.5~\% for the 4 bar case. Using a selection of events based on risetime we are able to remove edge effects and these values improve to $\sim$2\% FWHM (being the best value 1.8\% FWHM for the 4 bar case). This is, to our knowledge, the best energy resolution for alpha particles obtained with gas amplification in pure Xenon, and in particular better than the requirements set by \cite{Avignone:2007fu} for double beta decay searches. This result therefore experimentally demonstrates our claim that state-of-the-art Micromegas of the microbulk type are a feasible and competitive option for a Xe TPC for double beta decay.

We are currently focused on extending the present results to higher pressures, up to 10 bar. As future steps, we plan to test microbulk readouts with reduced amplification gap, which should provide better performance at higher pressures \cite{Giomataris:1998rc}, as well as to explore the effect of possible quenchers in Xenon on the energy resolution. Although the work here presented represents a basic step forward in demonstrating the absence of fundamental problems for Micromegas operation in pure Xe, more work is needed to deal with also important issues like long term stability, measurement with actual electron tracks, use of a pixelised readout to do tracking, and simultaneous measurement of energy and tracking. Work is ongoing also in this directions, under the framework of the NEXT collaboration. We are setting up right now a relatively large Micromegas TPC of dimensions 30 cm diameter and 35 cm drift of fiducial volume, able to hold $\sim$1 kg of Xe at 10 bar. In this fiducial volume high energy electron tracks are fully contained, and the previous issues can be thus studied.

\section{Acknowledgements}

We want to thank our colleagues of the group of the University of
Zaragoza and CEA/Saclay, as well as the colleagues from the NEXT
and RD-51 collaborations for helpful discussions and
encouragement, and specially A. Gimeno, D. Fortu\~no, J. Castel
and A. Ortiz from Zaragoza and J. P. Mols from Saclay for their
technical contributions to this work. We thank also R. de Oliveira
and his team at CERN for the manufacturing of the microbulk
readouts. We acknowledge support from the Spanish Ministry of
Science and Innovation (MICINN) under contract ref. FPA2008-03456,
as well as under the CUP project ref. CSD2008-00037 and the CPAN
project ref. CSD2007-00042 from the Consolider-Ingenio2010 program
of the MICINN. Part of these grants are funded by the European
Regional Development Fund (ERDF/FEDER). We also acknowledge
support from the European Commission under the European Research
Council T-REX Starting Grant ref. ERC-2009-StG-240054 of the IDEAS
program of the 7th EU Framework Program. Finally, we also
acknowledge support from the Regional Government of Arag\'{o}n
under contract PI001/08.

\section*{References} 



\end{document}